# DarkSPARC: Dark-Blood Spectral Self-Calibrated Reconstruction of 3D Left Atrial LGE MRI for Post-Ablation Scar Imaging

Mohammed S. M. Elbaz[1]

[1]Department of Radiology, Northwestern University Feinberg School of Medicine, Chicago, IL, USA

Corresponding Author:

Mohammed S. M. Elbaz, PhD, FSCMR

Department of Radiology

Northwestern University Feinberg School of Medicine

737 N. Michigan Ave Suite 1600

Chicago, IL, USA

Email: Mohammed.elbaz@northwestern.edu

Word count: 4997

## ABSTRACT

**Purpose:**

To develop DarkSPARC, a retrospective, training-free, self-calibrated spectral reconstruction method that converts routine bright-blood 3D left atrial (LA) late gadolinium enhancement (LGE) MRI into a dark-blood image, and to quantify its impact on LA scar–pool CNR, SNR, effective CNR (eCNR), and scar quantification accuracy.

**Methods:**

DarkSPARC embeds bright-blood LA LGE into a calibrator-conditioned $(N+1)$-dimensional spectral domain and reconstructs a dark-blood–like image using scan-specific spectral landmarks. A scan-specific 3D numerical phantom framework was built from LAScarQS post-ablation LGE by cloning remote myocardium into the LA wall and imposing controlled scar burden. Five baseline cases spanning the 5th–95th percentiles of native scar–pool CNR, each with multiple scar burdens and 10 CNR degradation levels, yielded 200 phantoms. For every phantom, LA scar–pool CNR, SNR, eCNR, and Scar% were measured on bright-blood and DarkSPARC images. In vivo performance was evaluated in 60 public post-ablation scans of atrial fibrillation patients.

**Results:**

In scan-specific phantoms, DarkSPARC increased LA scar–pool CNR, SNR, and eCNR over bright-blood in all 200 experiments, with DarkSPARC/bright-blood ratios up to ~30-fold for CNR and ~6-fold for SNR in lowest-CNR conditions. At 70% CNR degradation, bright-blood underestimated ground-truth LA Scar% by −37% to −54%, whereas DarkSPARC reduced bias to ~−3% to −5%. In vivo, DarkSPARC similarly improved metrics: median scar–pool CNR, SNR, and eCNR increased from 20.0 to 135.9 (~6.8×), 70.6 to 200.6 (~2.8×), 0.22 to 0.75 (~3.4×), respectively (all p<0.001), LA Scar% from 3.9% to 9.75%.

**Conclusion:**

DarkSPARC is a scan-specific, training-free reconstruction yielding dark-blood 3D LA LGE, boosting CNR/SNR/eCNR and stabilizing scar quantification without extra scans.

## INTRODUCTION

Late gadolinium enhancement (LGE) MRI of the left atrium (LA) is widely used to visualize post-ablation scar and residual fibrosis in atrial fibrillation (AF), and the extent and distribution of LA scar have been associated with arrhythmia recurrence after catheter ablation (1) (2,3) Standard 3D LGE MRI protocols generally use 3D bright-blood (BB) inversion-recovery sequences; however, the thin LA wall lies adjacent to a high-signal blood pool, and scar and blood therefore occupy overlapping intensity ranges. As a result, effective scar–pool contrast-to-noise ratio (CNR) is often low and fibrosis quantification can vary substantially across thresholds, readers, and software implementations.(3)

Sequence-level dark- or gray-blood LGE techniques can attenuate blood signal and improve scar conspicuity, but they require dedicated acquisitions in addition to BB LGE, increase scan time and sensitivity to timing and motion, and often exchange contrast gains for lower SNR. (4) (5) (6) (7) Deep-learning image-to-image translation approaches have been explored to synthesize alternative LGE contrasts, but they depend on paired or high-quality target data and may hallucinate or distort subtle enhancement when robust dark-blood ground truth is limited. (3)

To overcome these limitations, we propose DarkSPARC (Dark-blood SPectral self-calibrAted ReconstruCtion), a fully retrospective, training-free spectral reconstruction technique that converts standard bright-blood 3D LA LGE into a dark-blood–like image without additional acquisitions. DarkSPARC re-expresses the native $N$-dimensional image in a scan-specific, calibrator-conditioned spectral domain, using the LA blood signal as a statistical calibrator. The image is embedded into this domain as an $(N+1)$-dimensional spectral volume, where the added axis indexes standardized spectral bins encoding signal relative to the calibrator, rather than in arbitrary scanner intensity units. Intrinsic properties of the calibrator distribution define spectral bands corresponding to background, blood-pool–like, and supra-calibrator (scar-like) components. These bands are recombined to suppress blood and noise while preserving LA wall enhancement. The resulting DarkSPARC image maintains the same spatial resolution as the input LGE and is fully self-calibrated to the scan's own contrast and noise statistics.

A second challenge in this area is the lack of phantoms that combine true ground-truth LA scar with the heterogeneous, non-Gaussian statistics of clinical 3D LGE. Numerical CMR phantoms such as MRXCAT (8) provide realistic anatomy and motion but are not tailored to scan-specific LA LGE or post-ablation scar patterns, and thus cannot capture patient-specific texture, noise, and intensity inhomogeneity. To address this, a second contribution of this work is a scan-specific numerical phantom framework that builds 3D LGE phantoms by cloning each subject's own remote myocardium and scar into the LA wall, preserving native image statistics while enabling controlled, ground-truth LA scar burden for rigorous reconstruction testing.

Accordingly, the aims of this study were to: 1) describe the DarkSPARC spectral reconstruction technique for 3D LA LGE; 2) evaluate its accuracy and robustness to CNR loss in scan-specific numerical phantoms with known scar burden; and 3) assess its in vivo performance in post-ablation patients by quantifying improvements in CNR, SNR, and scar conspicuity relative to standard 3D bright-blood 3D LA LGE.

## METHODS

### Dataset

This study used the public LAScarQS 2022 (Left Atrial Scar Quantification & Segmentation) Task 1 dataset, which includes 60 post-ablation 3D LA LGE scans from three centers (University of Utah, Beth Israel Deaconess Medical Center, King's College London)(9). Center 1 scans were acquired on Siemens Avanto 1.5T and Verio 3T systems using free-breathing, navigator-gated 3D LGE (0.625 × 0.625 × 2.5 mm) at 3–6 months post-ablation. Centers 2 and 3 used Philips Achieva 1.5T systems with free-breathing, navigator-gated, fat-suppressed 3D LGE at 1 month (1.4 × 1.4 × 1.4 mm) and 3–6 months (1.3 × 1.3 × 4.0 mm), respectively. For all cases, the dataset provides 3D LA LGE plus binary LA cavity and scar masks, which were used to define the LA wall, native scar geometry, and seed ROIs for scan-specific scar cloning in the phantom experiments.

**The DarkSPARC Technique**

Bright-blood LA LGE is fundamentally ambiguous because blood pool and scar share overlapping intensity ranges. DarkSPARC (Dark-blood SPectral self-calibrAted ReconstruCtion) addresses this by re-expressing the image in a scan-specific, calibrator-conditioned spectral domain using LA blood as calibrator. The entire image is embedded into calibrator space (here blood signal), yielding an $(N+1)$-dimensional spectral volume in which each voxel is represented by a standardized probability density function (PDF) relative to the calibrator, and the additional axis indexes spectral bins of calibrator-relative signal.

From the calibrator statistics (e.g., its cumulative distribution function), DarkSPARC defines spectral bands that separate background, blood-pool–like, and supra-calibrator (scar-like) components. A dark-blood image is then reconstructed by appropriately combining these bands to suppress background and blood-related signal while preserving anatomy and scar morphology. The resulting DarkSPARC image has the same dimensions as the input bright-blood LGE, is fully self-calibrated and training-free, and derives all spectral axes and landmarks directly from the calibrator of the same scan. The technique's two-stages are detailed below and illustrated in Figure 1.

*Calibrator ROI $C$ Definition*

DarkSPARC is self-calibrated from a small "calibrator" region of interest $\mathbf{C}$ that samples the signal to be suppressed. Here, the left atrial blood pool is used as a calibrator. The method does not require full blood-pool segmentation: it only needs a rough ROI, which can be defined on a single representative slice rather than across the entire 3D volume. The role of this ROI is to provide a sufficient sample of calibrator voxels from which the spectral profile and its uncertainty are estimated (see section on sensitivity analysis to ROI choice below).

The calibrator (blood pool ROI) vector

$$\mathbf{C} = [g_j], j = 1, \dots, M \qquad (1)$$

was then defined as the set of voxel intensities within this blood-pool ROI. These $M$ samples are used to construct the calibrating spectral profile and its associated uncertainty in the subsequent spectral encoding stage.

*Stage 1: Spectral Probabilistic Profile Encoding*

In Stage 1, DarkSPARC constructs the calibrator-conditioned spectral domain by generating two probabilistic spectral profiles from the image and calibrator signals (Fig. 1). Spectral encoding in DarkSPARC was implemented using our recent probabilistic LGE signature framework (10).

The algorithm operates on the full bright-blood image intensity vector

$$\mathbf{I} = [g_i], i = 1, \dots, N \text{ voxels.} \qquad (2)$$

To re-encode the data in a calibrator-conditioned space, two disparity vectors are computed. The cross-disparity vector $\Phi_{\text{cross}}$ maps the entire image into the calibrator space via all pairwise products between $\mathrm{I}$ and $\mathrm{C}$:

$$\Phi_{\text{cross}(I,C)} = [g_i \cdot g_j], 1 \le i \le N,\ 1 \le j \le M. \qquad (3)$$

The calibrating auto-disparity vector $\Phi_{\text{auto-calib}}$ characterizes the intrinsic behavior of the calibrator within the same space, using pairwise products within $\mathrm{C}$:

$$\Phi_{\text{auto_calib}(C,C)} = [g_j \cdot g_l], 1 \le j \le M, j \le l \le M. \qquad (4)$$

The statistical distributions of these high-dimensional disparity vectors are then summarized as normalized histograms with $B$ bins, yielding two probability density functions (PDFs): the Image Spectral Profile $S_I$ and the Calibrating Spectral Profile $S_C$:

$$S_I = \text{pdf}(\Phi_{\text{cross}}, B), \qquad (5)$$
$$S_C = \text{pdf}(\Phi_{\text{auto-calib}}, B), \qquad (6)$$

with $B = 100$ bins in this implementation. Here, $S_I$ describes how the entire image co-distributes with the calibrator, whereas $S_C$ encodes the spectral “fingerprint” and uncertainty of the calibrator itself.

Scan-specific calibration landmarks $(\beta_0, \beta_1, \beta_2)$ are then derived from $S_C$ and its cumulative density function $\text{CDF}_{S_C}$:

$$\beta_0 = \min\{\beta: S_C(\beta) > 0\}, \quad (7)$$
$$\beta_1 = \arg\max_{\beta} S_C(\beta), \quad (8)$$
$$\beta_2 = \min\{\beta: \mathrm{CDF}_{S_C}(\beta) \geq 1 - \varepsilon\}. \quad (9)$$

Here, $\beta_0$is the first spectral bin with nonzero calibrator density, so bins $\beta < \beta_0$correspond to background or sub-calibrator noise. $\beta_1$is the mode of $S_C$and represents the most probable pure calibrator response (core of the calibrator spectrum). $\beta_2$marks the practical end of the calibrator distribution; bins $\beta > \beta_2$are supra-calibrator enhancements corresponding to scar and other highly enhanced structures. Conceptually, this endpoint is where the calibrator CDF reaches 1; in practice, we set $\beta_2$as the smallest bin index with $\mathrm{CDF}_{S_C}(\beta_2) \geq 1 - \varepsilon$, so at most an $\varepsilon$-fraction of calibrator mass lies beyond $\beta_2$, avoiding overfitting the uncertainty-dominated tail. Together, $(\beta_0, \beta_1, \beta_2)$delineate sub-calibrator, core-calibrator, and supra-calibrator regimes used as scan-specific integration bounds in Stage 2 (Fig 1).

*Stage 2: Self-Calibrated Spectral-Resolved Reconstruction*

Stage 2 reconstructs the spectral dimension, forming an $(N + 1)$-dimensional representation of the image. For an input 3D LGE volume $I(x, y, z)$of size $X \times Y \times Z$, DarkSPARC generates $B$bin-specific spatial maps (here $B = 100$) along the spectral axis, yielding a 4D spectral-resolved volume $\phi(x, y, z, \beta_r)$, $\beta_r \in \{1, \dots, B\}$. Each slice along this spectral axis is typically sparse and encodes a subset of the original image content grouped by calibrator disparity. This 4D volume is then used to compute four cumulative spectral maps by aggregating bin-specific contributions back into four $N$-dimensional components (Fig. 1). For each bin $\beta_r$, a disparity interval $[\beta_{\text{low}}, \beta_{\text{high}}]$is defined, and contributing voxels from the original image are identified by retracing the cross-disparity products using an indicator function $\delta$ as follows:

$$\delta_{\beta_{\text{low}},\beta_{\text{high}}}(\Phi_{\text{cross}}(g_i \cdot g_j)) = \begin{cases} 1, & \Phi_{\text{cross}}(g_i \cdot g_j) \in [\beta_{\text{low}}, \beta_{\text{high}}], \\ 0, & \text{otherwise.} \end{cases} \quad (10)$$

This selects all image–calibrator pairs whose disparity products fall within the current bin.

From these indices, a unique set of contributing voxel indices $U$ is formed:

$$U = \{i: i \text{ appears in the bin range (without duplicates)}\}. \quad (11)$$

For each unique voxel $u \in U$, its probabilistic contribution $h(u)$ to the current bin is computed as a normalized frequency, representing the fraction of its image–calibrator pairings that fall within the bin relative to all pairings assigned to that bin:

$$h(u) = \frac{\sum_{j=1}^{M} \delta_{\beta_{\text{low}},\beta_{\text{high}}}(\Phi_{\text{cross}}(g_u \cdot g_j))}{\sum_{i=1}^{N} \sum_{j=1}^{M} \delta_{\beta_{\text{low}},\beta_{\text{high}}}(\Phi_{\text{cross}}(g_i \cdot g_j))}. \qquad (12)$$

These voxelwise contributions are then assembled into the spectral-resolved dataset $\phi$. For a 3D input volume $I(x, y, z)$, this yields

$$\phi(x, y, z, \beta_r) = h(u), \qquad (13)$$

so that each slice $\phi(\cdot,\cdot,\cdot, \beta_r)$ is the bin-specific spatial map for spectral bin $\beta_r$. Stacking the $B$ bin-specific maps along $\beta_r$forms the full 4D spectral-resolved volume $\phi(x, y, z, \beta_r)$, i.e., the $(N + 1)$-dimensional representation of the image in the spectral calibrator-conditioned domain.

The spectral-resolved volume is then projected back into an $N$-dimensional spatial representation by integrating along the spectral axis using the calibration landmarks $(\beta_0, \beta_1, \beta_2)$ defined in Eqns. 7-9 as integration limits. This yields four cumulative spectral maps:

$$\phi_1(x, y, z) = \sum_{q=1}^{\beta_0} \phi(x, y, z, \beta_q), \qquad (14)$$

$$\phi_2(x, y, z) = \sum_{v=\beta_0+1}^{\beta_1} \phi(x, y, z, \beta_v), \qquad (15)$$

$$\phi_3(x, y, z) = \sum_{r=\beta_1+1}^{\beta_2} \phi(x, y, z, \beta_r), \qquad (16)$$

$$\phi_4(x, y, z) = \sum_{k=\beta_2+1}^{B} \phi(x, y, z, \beta_k). \qquad (17)$$

In this construction, $\phi_1$predominantly accumulates background and very low-signal voxels below the onset of the calibrator distribution; $\phi_2$ aggregates voxels whose spectral contributions follow the lower portion of the calibrator (calibrator-like) distribution; $\phi_3$captures the core and upper part of the calibrator distribution corresponding to the main bright-calibrator (bright blood) component; and $\phi_4$ accumulates supra-calibrator high-signal voxels, including scar and other structures whose disparity products extend beyond the main calibrator distribution (Fig. 1).

The final DarkSPARC dark-blood image $\Psi$ is obtained by combining these four component maps via a dual-contrast formulation (Fig. 1):

$$\Psi(x, y, z) = [\phi_2(x, y, z) - \phi_1(x, y, z)] + [\phi_4(x, y, z) - \phi_3(x, y, z)]. \tag{18}$$

The two difference terms provide complementary contrasts. The first, $\phi_2 - \phi_1$, suppresses background and low-signal fluctuations encoded in $\phi_1$while preserving structural detail in the lower calibrator-related band $\phi_2$. The second, $\phi_4 - \phi_3$, yields dark-blood contrast by suppressing the main bright-calibrator component in $\phi_3$and retaining supra-calibrator signal in $\phi_4$, including scar and other high-intensity enhancements. Their sum produces an $N$-dimensional image with reduced background noise, nulled calibrator compartment (blood pool in this study), and accentuated supra-calibrator enhancement. The final DarkSPARC image $\Psi$ matches the spatial dimensions of the input $I$ and is shifted to start at zero and linearly scaled to a 12-bit range (0–4095) for display and quantitative analysis (Fig 1).

**Hybrid Scan-Specific 3D LGE Numerical Phantom**

To rigorously validate DarkSPARC, we developed a scan-specific 3D LGE numerical phantom framework. Unlike idealized phantoms with simplified geometries and Gaussian noise, this approach preserves the complex, non-Gaussian statistical texture of the original clinical data while providing a fully traceable ground truth (Figs 2, 3). Phantom construction uses two deterministic stages: (1) normal-tissue cloning to create a pristine normal wall, and (2) scar cloning to impose the pathological ground truth.

*Stage 1: Normal Tissue Cloning*

The goal of this stage is to replace all non-scar myocardial intensities with a statistically “pure” normal tissue distribution derived from the patient’s own remote myocardium. This removes potential latent fibrosis and artifacts within the original wall, ensuring that the ground-truth healthy tissue is uncontaminated (Fig. 2).

Let $F(x)$ denote the empirical cumulative distribution function (CDF) of intensities in a reference region of interest (ROI), and $F^{-1}(p)$ its empirical inverse CDF (quantile) for $p \in [0,1]$. If $X$ is the set of intensity values in the reference remote ROI, the empirical CDF is

$$F(x) = P(X \leq x),$$

and the inverse CDF is implemented by sorting $X$ and linearly interpolating between order statistics. The cloning process follows the following steps:

1. *Reference statistics.* A scan-specific remote normal tissue reference distribution $F_{\mathrm{remote}}$ is computed from a manually defined ROI in the remote myocardium $M_{\mathrm{remote}}$.
2. *Spatial rank preservation.* To preserve the native spatial texture (e.g., coil-sensitivity gradients), the original intensity values within the entire left atrial wall mask $M_{\mathrm{wall}}$are sorted to determine the spatial rank of each voxel. These ranks are converted to quantiles $q_{\mathrm{wall}}(x) \in [0,1]$ for every voxel coordinate $x$.
3. *Intensity mapping.* The wall is re-synthesized by mapping these spatial ranks through the remote inverse CDF:

$$I_{\mathrm{cloned}}(x) = F_{\mathrm{remote}}^{-1}(q_{\mathrm{wall}}(x)), \forall x \in M_{\mathrm{wall}}.$$

Here, $F_{\mathrm{remote}}^{-1}(u) = \inf\{t: F_{\mathrm{remote}}(t) \geq u\}$. This operation statistically “clones” the remote tissue distribution onto the entire wall (i.e. zero scar) while preserving the original spatial structure (Fig. 2).

<u>*Stage 2: Scar Cloning and Modulation*</u>

After normalizing the wall with cloned normal tissue, the scar component is imposed. To ensure quantitative traceability ground-truth scar, a scar-floor threshold

$T = \mu_{\mathrm{pool}} + 3.3\,\sigma_{\mathrm{pool}}$

is enforced which was previously histologically-validated (11) (Fig. 3), where $\mu_{\mathrm{pool}}$and $\sigma_{\mathrm{pool}}$are the mean and standard deviation of the blood-pool intensities. Scar is integrated into the phantom in two modes:

1. *Original scar retention (native phantoms).* For “native” phantoms, the original intensities from the baseline scan are restored within the original manual scar mask $M_{\text{scar}}$ (provided in the dataset) and floored at $T$:

$$A(x) = \max\left(I_{\text{original}}(x), T\right), \forall x \in M_{\text{scar}}.$$

2. *Modulated scar cloning (10–30% phantoms).* To achieve specific scar burdens (e.g., 10%, 20%, 30% of LA wall volume), new scar voxels are added contiguously to the original scar boundary using a distance-weighted priority map based on Euclidean distance transform. To maintain realism, their intensities are cloned from the patient’s actual scar distribution similar to the normal tissue cloning above but this time for scar. A reference scar distribution $F_{\text{scar}}$is computed from $M_{\text{scar}}$. New intensities are synthesized by sampling the upper tail of this distribution (intensities $> T$) via inverse CDF mapping:

$$I_{\text{new}}(x) = F_{\text{scar}}^{-1}\left(q_{\text{tail}}(x)\right),$$

where $q_{\text{tail}}$are quantiles re-mapped to the upper probability range of the scar distribution. This yields contiguous, realistic scar patterns at prescribed global scar burdens (Fig. 2).

*Code for the scan-specific numerical phantom framework will be made publicly available upon publication.*

**Phantom Validation Experimental Design**

*Baseline Case Selection*

Five clinical LA LGE datasets were selected from the LAScarQS cohort to serve as baselines. To represent the full spectrum of clinical image quality, these cases corresponded to the 5th, 25th, 50th, 75th, and 95th percentiles of the cohort’s baseline scar-to–blood-pool contrast-to-noise ratio (CNR).

*Scar Burden Cohorts*

For each of the 5 baseline cases, four ground-truth phantoms were constructed by modulating the scar burden:

1. *Native phantom:* cloned normal tissue + original scar percentage (from manual segmentation).
2. *10% phantom:* cloned normal tissue + scar cloned to 10% of LA wall volume.
3. *20% phantom:* cloned normal tissue + scar cloned to 20% of LA wall volume.
4. *30% phantom:* cloned normal tissue + scar cloned to 30% of LA wall volume.

This yielded 20 unique ground-truth phantoms ($5$ baselines $\times$ $4$ burdens).

<u>*Sensitivity to CNR Degradation*</u>

To assess robustness to noise and contrast loss, each of the 20 ground-truth phantoms was subjected to controlled CNR degradation. Rician noise was injected to simulate a linear fractional decline in absolute CNR. Simulations were performed at 9 degradation levels (10–90% CNR reduction in 10% steps), plus the non-degraded baseline. The additional noise standard deviation was

$$\sigma_{\text{add}} = \sqrt{\sigma_{\text{tgt}}^2 - \sigma_0^2}, \sigma_{\text{tgt}} = \frac{\sigma_0}{1-d},$$

where $\sigma_0$is the original noise standard deviation derived from air background ROI and $d$ is the desired fractional CNR reduction.

This design generated 200 phantom experiments ($5$ baselines $\times$ $4$burdens $\times$ $10$degradation levels). For each experiment, quantification accuracy (relative scar error) and image quality (absolute CNR, SNR, and effective CNR [eCNR]) were evaluated for both the original bright-blood and DarkSPARC reconstructions.

**Phantom Quantitative Analysis**

For each phantom volume (all scar configurations and CNR levels), DarkSPARC was applied to the degraded bright-blood image to generate a dark-blood reconstruction with identical geometry. ROI masks for LA blood pool, LA scar, LA wall, and background air were inherited from the phantom construction and used identically for bright-blood and DarkSPARC.

Noise standard deviation was estimated from the air ROI. For each reconstruction, we computed:

- SNR in each ROI as mean signal divided by noise standard deviation.
- Scar–pool CNR as $(\mu_{\text{scar}} - \mu_{\text{pool}})/\sigma_{\text{noise}}$.
- Effective CNR (eCNR) as

$$\text{eCNR} = \frac{\mu_{\text{scar}} - \mu_{\text{pool}}}{\sqrt{\sigma_{\text{scar}}^2 + \sigma_{\text{pool}}^2}},$$

using a pooled standard deviation to capture intrinsic separation between scar and blood intensity distributions rather than pure noise scaling, consistent with prior eCNR-based evaluations. (12)

LA Scar% on bright-blood and DarkSPARC images was defined as the percentage of LA wall voxels with intensity $\geq T = \mu_{\text{pool}} + 3.3\,\sigma_{\text{pool}}$, using the same blood-pool–based threshold as in phantom construction. Ground-truth Scar% was taken from the scar masks for each phantom (native, 10%, 20%, 30%). For each phantom condition and CNR level, relative Scar% error for bright-blood and DarkSPARC was computed as the percent difference between measured Scar% and ground truth. Errors and image metrics (SNR, CNR, eCNR) were then summarized across baseline phantoms, scar burdens, and CNR levels to compare bright-blood versus DarkSPARC performance.

**Calibrator ROI sensitivity analysis**

We evaluated DarkSPARC's sensitivity to calibrator ROI placement and size in the native baseline phantom with the smallest scar burden (the most challenging, calibration-sensitive case), across all five baselines and nine CNR degradation levels (50 phantoms). The LA blood-pool ROI was perturbed in three ways: (i) changing slice location from the nominal mid-area (50% LA area) slice to a more basal 25%-area slice, (ii) changing ROI size at the 50%-area slice (larger erosion 9 vs smaller erosion 17), and (iii) jointly changing both location and size (75%-area slice with erosion 17 vs 50%-area slice with erosion 9). For each perturbation, DarkSPARC was rerun across all degradation levels, and Scar% error, CNR, and SNR were recomputed.

**In vivo post-ablation 3D LGE evaluation**

DarkSPARC was also evaluated in vivo using the 60 post-ablation 3D LA LGE scans in the public LAScar / LAScarQS 2022 dataset described above. For each case, the provided bright-blood 3D LGE volume was used as input to DarkSPARC to generate a dark-blood reconstruction with identical coverage, voxel size, and orientation. The supplied LA segmentations in the dataset were used to define the LA wall mask and LA cavity (blood-pool) ROI, and a background air ROI was placed outside the thorax for noise estimation. The same masks and ROIs were applied identically to both bright-blood and DarkSPARC images.

For every subject, SNR, scar–to–blood-pool CNR, effective CNR (eCNR), were computed on both reconstructions using the same definitions as in the phantom experiments. In vivo LA Scar% was defined, for each method, as the percentage of LA wall voxels with intensity at or above a blood-pool–based threshold $T_{\text{in vivo}} = \mu_{\text{pool}} + 3.3\,\sigma_{\text{pool}}$, where $\mu_{\text{pool}}$and $\sigma_{\text{pool}}$were measured from the bright-blood LA blood-pool ROI. Unlike in the phantoms, in vivo Scar% represents an imaging-derived estimate without a definitive ground truth. All metrics (SNR, CNR, eCNR, and Scar%) were obtained twice per subject—once from bright-blood and once from DarkSPARC—enabling direct within-subject comparison between the two reconstructions.

**Statistical Analysis**

Quantitative phantom results were summarized as mean ± standard deviation across the five baseline CNR setups, scar-burden configurations, and CNR degradation levels, characterizing trends in CNR, eCNR, SNR, contrast-to-noise enhancement, and relative Scar% error for bright-blood and DarkSPARC. For the calibrator ROI sensitivity analysis, agreement between metrics from nominal and perturbed blood-pool ROIs was assessed using mean percentage difference and intraclass correlation coefficients (ICC) for CNR, SNR, and Scar%. For the in vivo LAScar cohort, continuous variables were reported as median [25th percentile, 75th percentile] because they violated normality on the Shapiro–Wilk test. Within-subject differences in CNR, eCNR, and SNR between bright-blood and DarkSPARC were evaluated using the paired Wilcoxon signed-rank test. A P value $< 0.05$ was considered statistically significant.

## RESULTS

### Baseline numerical phantoms and spectral behavior across CNR

Five baseline post-ablation 3D LA LGE phantoms were generated from LAScar cases spanning the 5th, 25th, 50th, 75th, and 95th percentiles of native bright-blood scar–pool CNR, while preserving the original anatomy, noise texture, and intensity inhomogeneity outside the LA wall (Fig. 2). For each baseline case, four phantoms were constructed (native Scar% plus 10%, 20%, and 30% scar-cloned phantoms), yielding 20 distinct scar configurations on realistic image backgrounds.

Across this CNR spectrum, DarkSPARC spectral decomposition showed a consistent progression of spectral content (Fig. 3). The lowest-index spectral bins predominantly captured background and very low-signal noise-like voxels; intermediate bins followed blood-pool–like intensities and textures; and higher-index bins increasingly emphasized supra-calibrator components, including LA scar and other high-intensity enhancement. The corresponding DarkSPARC reconstructions exhibited dark-blood suppression with preserved LA wall enhancement across both low- and high-CNR baselines (Figs. 3–5).

### Phantom scar quantification and CNR degradation

Across all baseline CNR levels, scar burdens, and degradation steps, DarkSPARC increased LA scar–pool CNR, SNR, and eCNR relative to bright-blood in every phantom experiment, with DarkSPARC/bright-blood CNR and eCNR gain factors >1.0 at all CNR settings and largest in the lowest-CNR regimes (Fig. 4, Fig. 5). Thus, in the phantoms, DarkSPARC not only preserved contrast as CNR declined but actively boosted scar–blood separability compared with the original bright-blood images.

At the same time, DarkSPARC markedly stabilized LA Scar% estimates across CNR degradations (Table 1, Fig. 5). For the native-scar phantoms (4.3 ± 1.4% Scar), DarkSPARC maintained near-zero bias at mild degradation (0.8 ± 1.9% at 10%; 1.2 ± 2.8% at 20%) and showed substantially smaller underestimation at higher degradation (−3.1 ± 8.2%, −14.3 ± 14.5%, and −26.3 ± 23.0% at 70%, 80%, and 90% degradation, respectively). In contrast, bright-blood LGE already underestimated Scar% by −19.4 ± 1.9% and −26.5 ± 16.8% at 10% and 20% degradation, worsening to −53.9 ± 24.6%, −66.6 ± 25.1%, and −84.5 ± 17.3% at 70%, 80%, and 90% degradation (Table 1, Fig. 4). Notably, bright-blood bias at only 20% degradation (−26.5 ± 16.8%) was already comparable in magnitude to DarkSPARC error at 90% degradation (−26.3 ± 23.0%).

The same behavior was observed in the 10%, 20%, and 30% scar-cloned phantoms (Table 1, Fig. 3). At 10% CNR degradation, DarkSPARC errors remained small (0.4–0.6%), whereas bright-blood underestimation ranged from −8.2 ± 4.4% (10% burden phantoms) to −3.9 ± 2.1% (30% burden phantoms). At 70% degradation, DarkSPARC errors were only −3.7 ± 5.0% to −4.5 ± 4.4%, compared with bright-blood errors of −37.0 ± 22.0% to −41.3 ± 22.5%. At 90% degradation, bright-blood errors were consistently severe (−80.0 ± 22.5% to −81.2 ± 21.0%), while DarkSPARC errors, although increased, remained substantially closer to ground truth (−38.8 ± 17.3% to −45.4 ± 20.7%). Together with the CNR and eCNR gains (Fig. 5), these results show that DarkSPARC simultaneously enhances contrast and resists CNR-driven deterioration in LA Scar% quantification.

**In vivo post-ablation performance**

In the 60 post-ablation 3D LA LGE scans, DarkSPARC consistently improved quantitative image quality metrics compared with BB when analyzed using identical LA segmentations and ROIs (Table 2, Fig. 6, Fig. 7). Representative in vivo examples across all quartiles of baseline BB CNR illustrate the same substantially improved LA blood suppression and more conspicuous LA wall scar on DarkSPARC compared with BB (Fig. 6). LA scar–pool CNR increased ~7-fold from 20.0 [11.1, 27.8] on BB to 135.9 [71.8, 263.4] on DarkSPARC ($p < 0.001$). SNR increased from 70.6 [55.1, 90.4] to 200.6 [114.1, 419.9] ($p < 0.001$; ≈2.8-fold gain), and eCNR increased from 0.22 [0.15, 0.33] to 0.75 [0.56, 0.84] ($p < 0.001$; ≈3.4-fold gain). These in vivo CNR/SNR/eCNR gains are consistent with the phantom results (Fig. 7).

Using the same blood-pool–based threshold ($\mu_{\text{pool}} + 3.3\sigma_{\text{pool}}$) for both reconstructions, LA Scar% was significantly higher on DarkSPARC than on BB (9.75 [5.2, 17.4]% vs 3.9 [1.7, 9.6]%, $p < 0.001$; Table 2, Fig. 7). This direction and magnitude align with the phantom experiments (Table 1, Fig. 6), where BB systematically underestimated true Scar% as CNR decreased, whereas DarkSPARC reduced this underestimation while improving CNR and eCNR.

### Sensitivity to calibrator ROI definition

Across all nine CNR phantom degradation levels, blood pool ROI (calibrator) perturbations had minimal impact on DarkSPARC metrics (Table 3). Mean signed percentage differences relative to the nominal ROI were 0.0–0.3% for Scar% error, −0.5–2.0% for CNR, and −0.1–0.0% for SNR, with intraclass correlation coefficients ≥ 0.98 for all metrics and configurations. This indicates that DarkSPARC performance is highly robust to reasonable variations in calibrator slice location and ROI size.

## DISCUSSION

### *Main findings*

This work introduces DarkSPARC, a retrospective, training-free spectral reconstruction that converts standard bright-blood 3D LA LGE into a dark-blood–like image with boosted CNR and SNR using a scan-specific LA blood signal as calibrator. It operates entirely on routinely acquired bright-blood data, requiring no additional scans, sequence changes, or external training, and is directly applicable to existing studies.

In scan-specific 3D LGE phantoms with known ground-truth scar, DarkSPARC consistently increased LA scar–pool CNR and eCNR across all CNR regimes, particularly at low CNR, while stabilizing Scar% quantification. Bright-blood underestimated LA Scar% by roughly −60% to −80% at the most severe CNR loss, whereas DarkSPARC kept bias within about ±1% up to 20% CNR loss, ~3–5% at 70%, and remained substantially less biased even at 80–90% degradation. In 60 post-ablation patients, median LA scar–pool CNR increased ~7-fold, SNR ~3-fold, and eCNR ~3.4-fold, using identical segmentations and ROIs. With the same blood-pool–based threshold ($\mu_{\text{pool}} + 3.3\sigma_{\text{pool}}$), LA Scar% rose from 3.9% (bright-blood) to 9.75% (DarkSPARC). Together, these phantom and in vivo results show that a scan-specific, training-free reconstruction can substantially improve CNR, SNR, and eCNR, enhance blood–scar separability, and reduce CNR-driven underestimation of LA scar compared with conventional bright-blood imaging.

*Relation to existing bright-blood, dark-blood, and AI-based approaches*

LA post-ablation scar imaging is fundamentally constrained by the thin LA wall, partial-volume effects, and overlapping scar–blood intensity distributions. In standard bright-blood workflows, methods that threshold directly in native intensity space become unstable as these distributions overlap: modest changes in CNR, noise, or coil sensitivity can drive large swings in measured Scar%, resulting in poor reproducibility and strong site- and protocol-dependence. (13) (14)

Sequence-based dark-blood LGE techniques (e.g., double inversion, magnetization transfer, motion- or flow-sensitized preparations) improve visualization by suppressing blood, but require modified sequences, precise timing, and are vulnerable to arrhythmia and heart-rate variability. (15) (16) (17) They add scan complexity, are not universally available across vendors, and cannot be applied retrospectively to existing bright-blood datasets limiting the potential for large studies. (15) (16) (17) Training-based AI or GAN-style image-to-image approaches face a different challenge: paired bright-blood/dark-blood ground truth is scarce, intensity statistics vary across vendors and protocols, and data-driven models may hallucinate or suppress subtle lesions in ways that are difficult to control, with performance tied to the training distribution.(18)

DarkSPARC takes a different route: it is a purely reconstruction-based, training-free method that operates directly on standard bright-blood acquisitions. By re-expressing the image in a scan-specific, calibrator-conditioned spectral domain, it can be applied prospectively and retrospectively without sequence modifications or external training data.

*Numerical Phantom framework and mechanistic insight*

A key contribution of this work is a scan-specific numerical phantom framework that bridges the gap between idealized digital phantoms and real patient datasets without ground truth. By cloning each patient's remote myocardium into the LA wall and imposing scar via a blood-pool–based intensity threshold, these phantoms preserve non-Gaussian texture, noise, and intensity inhomogeneity while providing verifiable ground-truth LA Scar%. Systematically varying scar burden on fixed anatomy, combined with CNR degradation across multiple baseline CNR levels, decouples scar burden from CNR and enables testing of reconstruction behavior under controlled yet clinically realistic conditions. Conventional numerical CMR phantoms such as MRXCAT, while anatomically realistic, are not tailored to scan-specific LA LGE or post-ablation scar statistics, underscoring the need for such complementary scan-specific phantoms. (19)

Within this framework, bright-blood Scar% error became increasingly negative as CNR declined, reaching ≈−55% to −66% at 80% degradation and ≈−80% at 90%. In contrast, DarkSPARC preserved the correct ordering and magnitude of 10–30% Scar%, with near-zero error at native and mild degradation, residual errors of only a few percent at 70%, and bias limited to ≈−15% to −17% at 80% despite severely degraded native contrast. These findings support a mechanistic view of DarkSPARC as a stabilizing reconstruction: by operating in a calibrator-conditioned spectral domain and suppressing blood-like components and background noise, it mitigates the bias that arises when scar and blood overlap in the native intensity space. Further work should also test whether the CNR and SNR gains can enable reduced gadolinium dose or shorter acquisitions without compromising diagnostic quality.

*In vivo interpretation and potential clinical implications*

The in vivo experiments extend the phantom findings to clinical post-ablation imaging. Using identical LA segmentations and the same $\mu_{\text{pool}} + 3.3\sigma_{\text{pool}}$ threshold on both reconstructions, DarkSPARC increased scar–pool CNR by ≈6.8×, SNR by ≈2.8×, and eCNR by ≈3.4×. Because eCNR incorporates pooled scar–blood variance, these gains indicate not only noise suppression but also greater intrinsic separation between scar and blood intensities.

The ≈2.5-fold higher LA Scar% on DarkSPARC aligns with phantom results, where bright-blood systematically underestimated Scar% at modest/low CNR while DarkSPARC remained unbiased at native or mildly degraded CNR. This likely reflects recovery of true scar signal previously submerged within overlapping blood intensity, rather than blooming. Clinically, this may be most relevant when LA LGE CNR is modest (early post-ablation, arrhythmia/tachycardia, lower field strength, or short scans), where bright-blood images often appear equivocal.[1–3] Because DarkSPARC is applied purely as post-processing, it can be used on new and existing LA LGE datasets without changing acquisition. This makes it a practical complement or alternative to sequence-based dark-blood imaging, especially in centers where such sequences are unavailable or not reliably implemented.(15) (16) (17) (20)

*Limitations*

This study has several limitations. First, the in vivo analysis was retrospective and limited to a specific multi-center LA LGE dataset acquired on two vendor platforms with fixed protocols and reconstruction settings; generalizability to other field strengths, vendors, and sequence implementations remains unknown.(13) Second, no prospectively acquired paired dark-blood LGE images were available, so comparisons were limited to conventional bright-blood LGE and did not directly benchmark against state-of-the-art sequence-based dark-blood techniques. (15) (16) (17) (20) Third, definitive in vivo ground truth for LA scar (e.g., histopathology or electroanatomic mapping) was not available, precluding direct validation of absolute scar burden in vivo. (14)

**CONCLUSIONS**

DarkSPARC retrospectively reconstructs dark-blood–like 3D LA LGE from conventional bright-blood acquisitions without extra scan time or training data, substantially boosting CNR, SNR. In scan-specific phantoms and 60 post-ablation patients, it reduced CNR-dependent underestimation of LA scar versus bright-blood LGE, supporting its potential as a more robust reconstruction for LA scar imaging from existing datasets.

**Acknowledgements:** This work was supported in part by the National Heart, Lung, and Blood Institute of the National Institutes of Health (R01HL169780).

**Declaration of competing interests:** Mohammed S. M. Elbaz is the inventor on a pending patent application, assigned to Northwestern University, related to the methods described in this manuscript.

## TABLES

**Table 1.** Mean relative scar error (mean ± SD, %) from numerical scar phantom experiments across the five baseline cases at key CNR degradation levels.
Negative values indicate underestimation of scar burden relative to ground truth.

| **Scar Burden (GT)** | **Method** | **10% degrd. (Error %)** | **20% degrd. (Error %)** | **70% degrd. (Error %)** | **80% degrd. (Error %)** | **90% degrd. (Error %)** |
|---|---|---|---|---|---|---|
| **Original/Native (4.3%±1.4% Scar)** | Bright-Blood | -19.4% ± 1.9% | -26.5% ± 16.8% | -53.9±24.6 | -66.6±25.1 | -84.5±17.3 |
| | DarkSPARC | 0.8±1.9 | 1.2±2.8 | -3.1±8.2 | -14.3±14.5 | -26.3±23.0 |
| **10% phantom** | Bright-Blood | -8.2±4.4 | -12.5±7.5 | -41.3 ± 22.5 | -57.8±27.6 | -81.2±21.0 |
| | DarkSPARC | 0.6±1.4 | 0.8±1.8 | -3.7±5.0 | -15.5±11.3 | -38.8±17.3 |
| **20% phantom** | Bright-Blood | -5.0±2.7 | -8.4±5.2 | -37.9±22.2 | -55.3±28.5 | -80.4±22.1 |
| | DarkSPARC | 0.4±1.0 | 0.5±1.2 | -4.4±4.5 | -16.7±11.6 | -43.5±19.4 |
| **30% phantom** | Bright-Blood | -3.9±2.1 | -7.0±4.4 | -37.0±22.0 | -54.6±28.6 | -80.0±22.5 |
| | DarkSPARC | 0.4±0.8 | 0.4±0.9 | -4.5±4.4 | -17.4±11.9 | -45.4±20.7 |

**Table 2:** In vivo Results over the 60 post-ablation cases

| Metric | Bright-Blood | DarkSPARC | p-value |
| --- | --- | --- | --- |
| **CNR** | 20.0 [11.1, 27.8] | 135.9 [71.8, 263.4] | <0.001 |
| **SNR** | 70.6 [55.1, 90.4] | 200.6 [114.1, 419.9] | <0.001 |
| **eCNR** | 0.22 [0.15, 0.33] | 0.75 [ 0.56 , 0.84] | <0.001 |
| **Scar%** | 3.9 [ 1.7, 9.6] | 9.75 [5.2, 17.4] | <0.001 |

**Table 3.** Effect of calibrator ROI perturbation on DarkSPARC metrics in the native baseline phantom with the smallest scar burden (most sensitive case).

| **Blood Pool ROI perturbation** | **Metric** | **Mean % difference (%)** | **ICC** |
|---|---|---|---|
| **ROI location: 25% vs 50% LA slice** | Scar% error | 0.0 | 0.98 |
| | CNR | 2.0 | 0.99 |
| | SNR | 0.0 | 1.00 |
| **ROI size: small vs large at 50% LA slice** | Scar% error | 0.0 | 0.98 |
| | CNR | 0.0 | 0.99 |
| | SNR | −0.1 | 1.00 |
| **ROI location + size: 75% vs 50% LA slice** | Scar% error | 0.3 | 0.98 |
| | CNR | −0.5 | 0.99 |
| | SNR | −0.1 | 1.00 |

## FIGURES

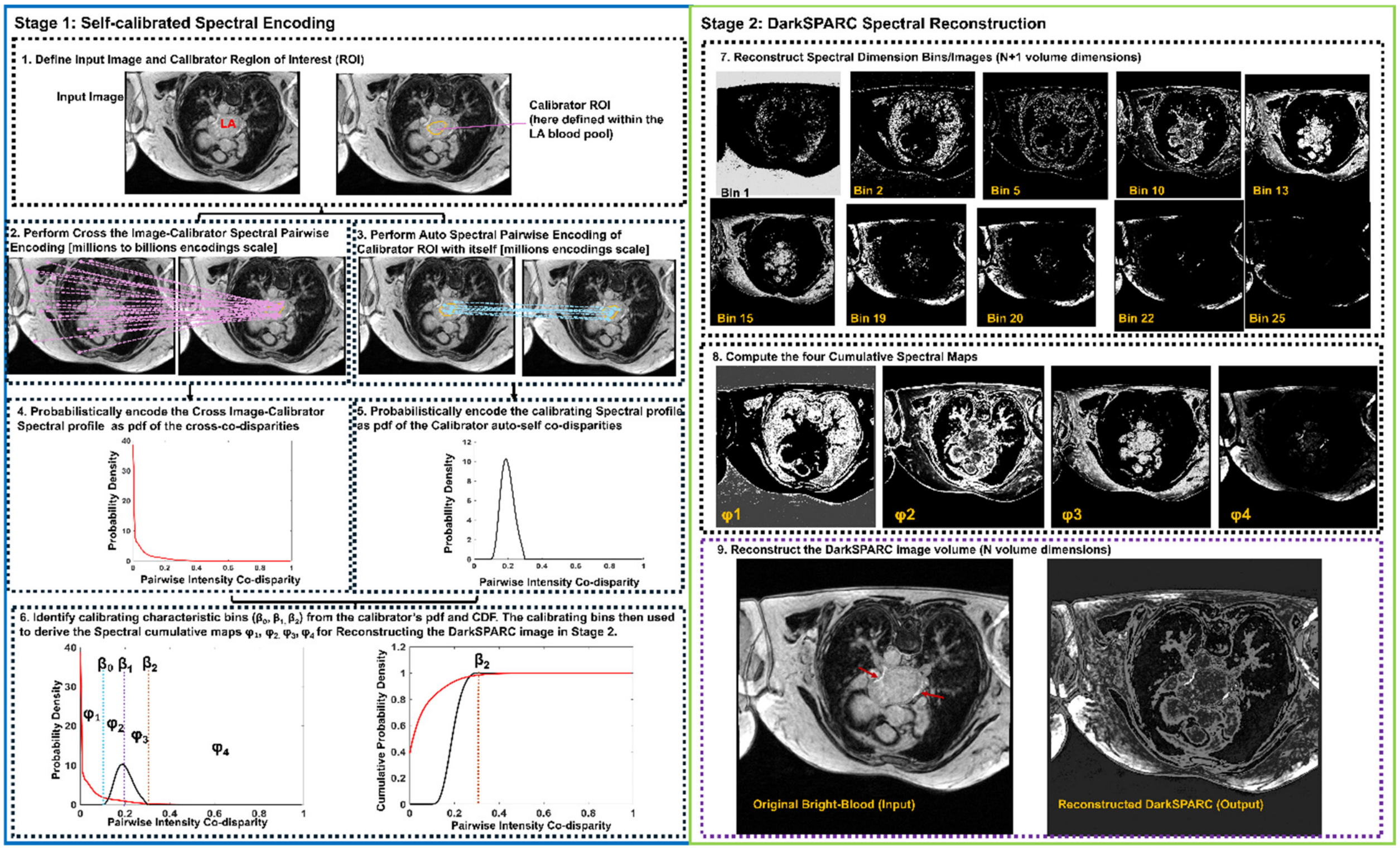

**Figure 1.** Overview of the proposed DarkSPARC reconstruction framework. The input is Bright-blood (BB) 3D LGE MRI acquired with standard clinical parameters. A rough left atrial (LA) blood-pool calibrator ROI is identified and used to define a scan-specific calibrator. In stage 1, The entire 3D LGE MRI image volume (not just LA) voxel intensities are then embedded/mapped into a calibrator-conditioned probabilistic spectral domain, through spectral pdf encoding of the cross-probability density function $\Phi_{cross}$, and spectral landmarks are defined from the calibrator profile. In stage, the newly calibrator-embedded image spectra are reconstructed into N+1 spectral dimension where each bin resolves sparse image details relative to the disparity from calibrator.  DarkSPARC reconstructs a dark-blood 3D LGE image volume by recombining scar-enhancing spectral components while suppressing blood-pool and background noise contributions, yielding a self-calibrated, dark-blood-like reconstruction that preserves scar contrast and anatomical detail.

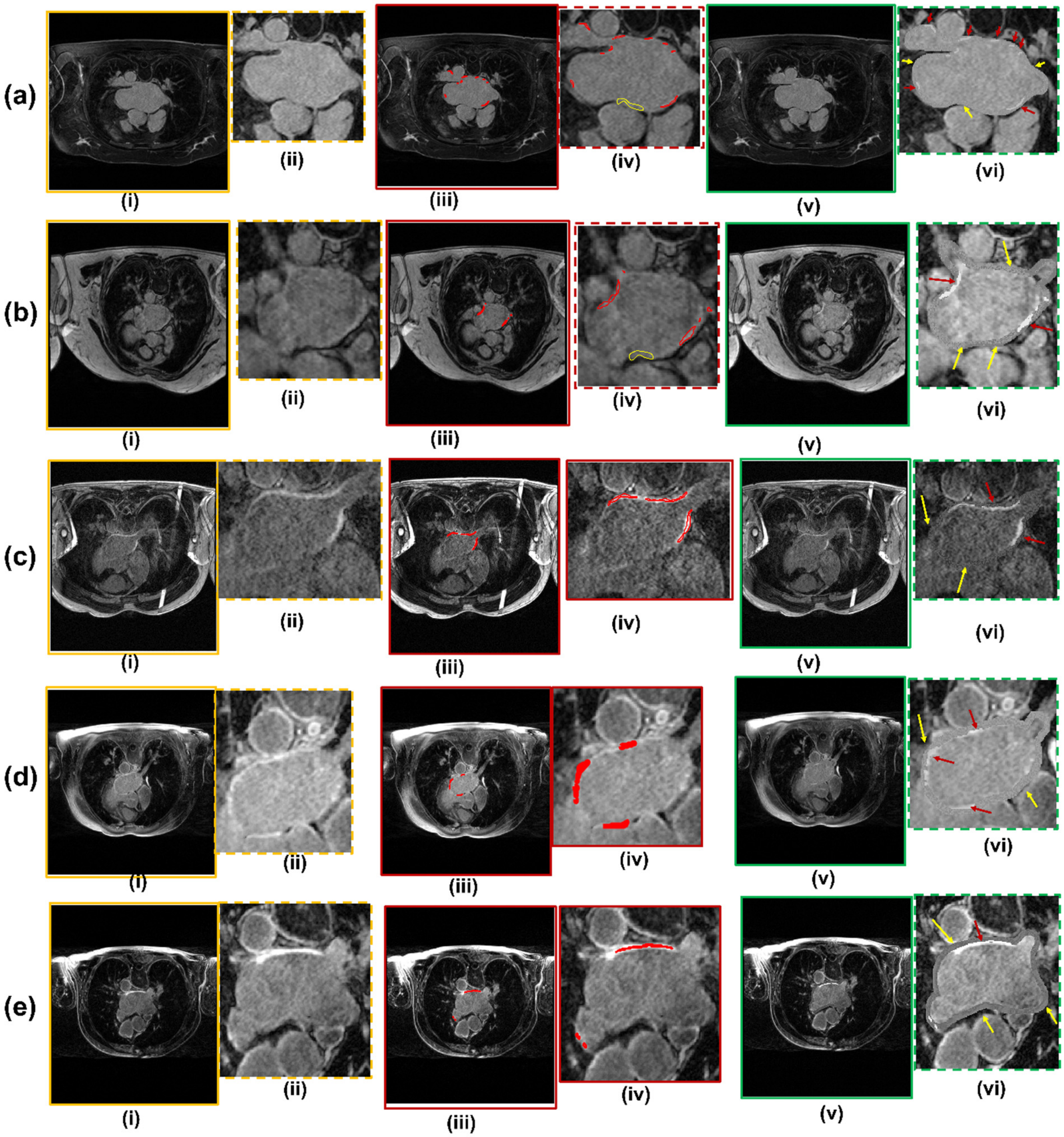

**Figure 2.** Construction of numerical baseline phantoms across the spectrum of in vivo scar–pool CNR. Five post-ablation LA LGE scans were selected to span the distribution of native bright-blood scar–pool CNR, approximating the 5th, 25th, 50th, 75th, and 95th percentiles (rows a–e, respectively) in the in vivo dataset. For each case, panels show cross-sections of the generated 3D phantoms of: (i) the original in vivo 3D bright-blood LGE image; (ii) a zoomed view of the LA region from (i); (iii) ground-truth scar ROIs (red contours) as provided by the dataset together with representative remote normal tissue examples (illustrated for rows a and b); (iv) a zoomed LA view of (iii); (v) the generated numerical phantom obtained by cloning remote normal tissue and scar within the LA wall, leaving all extra-atrial anatomy, noise texture, and intensity inhomogeneity unchanged, thereby preserving the native in vivo image distribution and uncertainty while enabling controlled wall-level ground-truth scar; and (vi) a zoomed LA view of (v), where red arrows indicate retained original scar regions and white arrows indicate regions where original wall intensities has been replaced by cloned normal tissue.

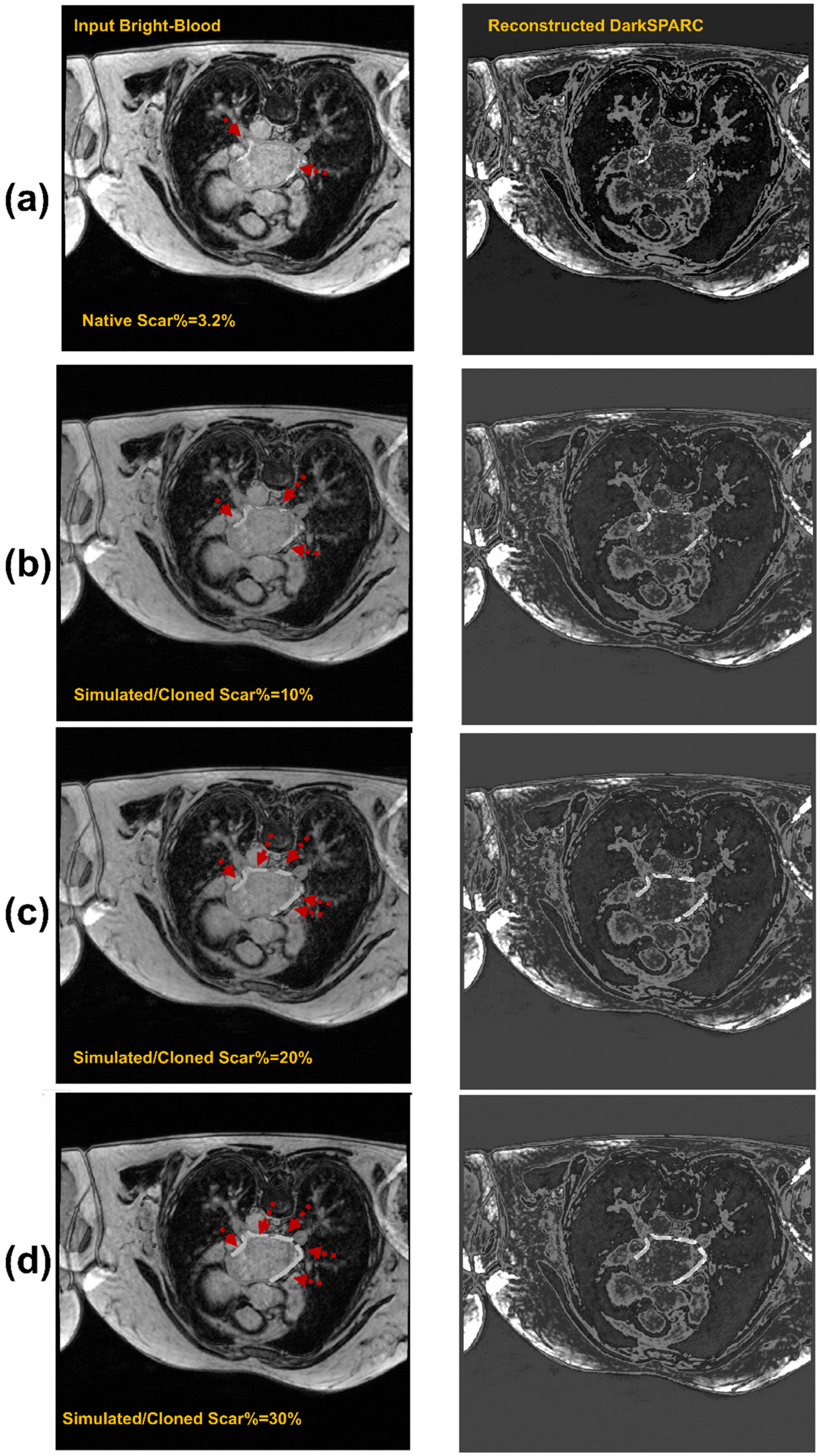

**Figure 3.** Example DarkSPARC spectral decomposition across four baseline phantom CNR levels showing example cross-sections of the 3D results.

Representative DarkSPARC results are shown for the 5th, 50th, 75th, and 95th percentile baseline CNR phantoms (rows a–d; the 25th percentile case is illustrated in Fig. 1 under the reconstruction schematic). For each phantom, panels include: (i) the original bright-blood LGE image; (ii) the corresponding DarkSPARC reconstruction; and (iii–vi) cumulative spectral maps associated with the four spectral bins $\Phi_1$–$\Phi_4$obtained in the calibrator-conditioned spectral domain. Lower-index bins primarily reflect background and sub-calibrator noise, intermediate bins progressively capture blood-pool-dominated and calibrator-like components, and higher-index bins correspond to supra-calibrator enhancement including scar. Across increasing native scar–pool CNR, DarkSPARC consistently suppresses LA blood-pool signal and background noise and increases scar conspicuity. Red arrows highlight representative bright-blood scar regions which can then be noted to their enhanced depiction on the DarkSPARC reconstructions.

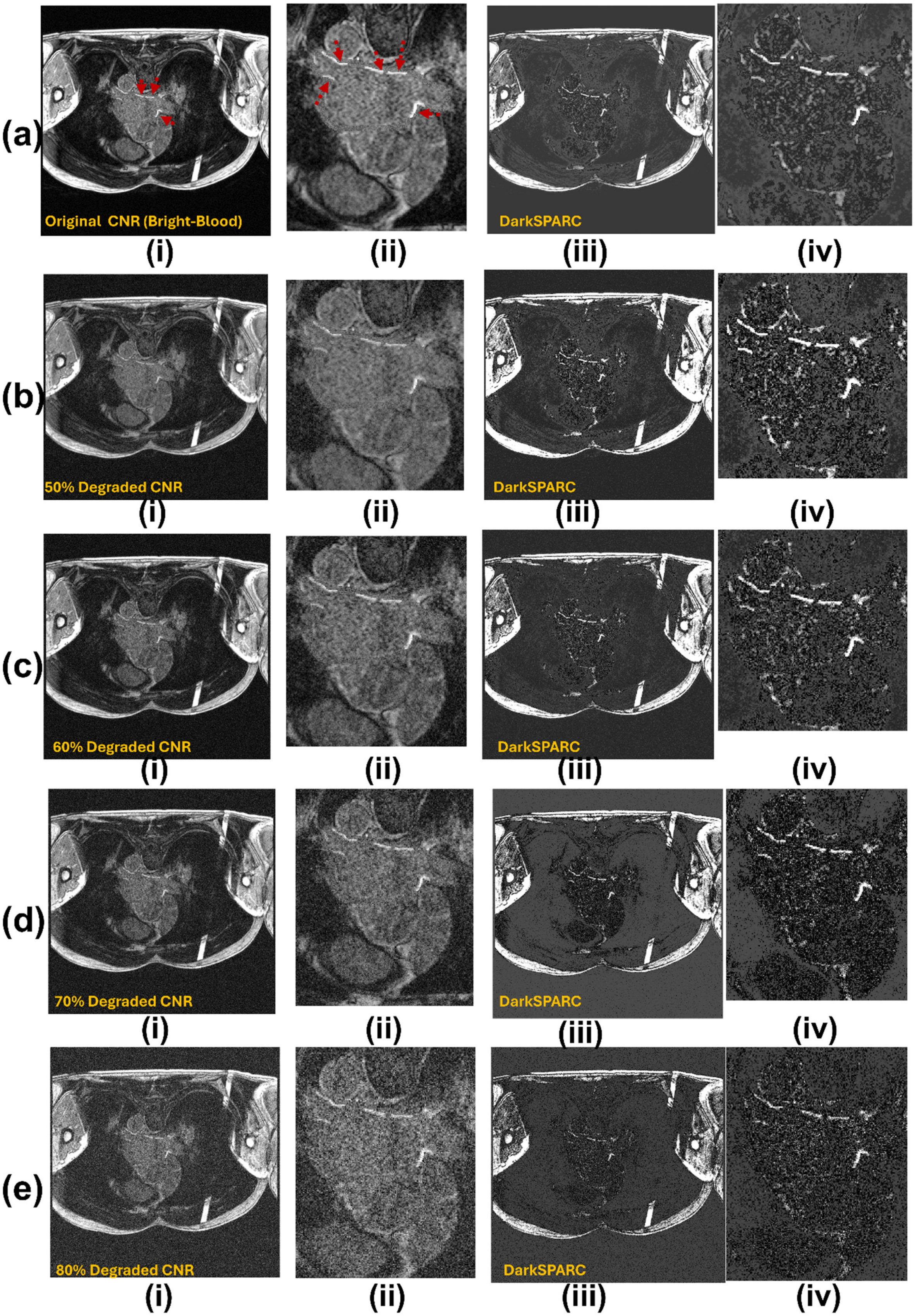

**Figure 4.** Effect of CNR degradation on bright-blood phantoms and corresponding DarkSPARC reconstructions. CNR degradation was applied to all five baseline phantoms; an example is shown for the median (50th percentile) baseline CNR phantom. Rows (a–e) show the same cross-section at (a) original CNR (no degradation), (b) 50 percent, (c) 60 percent, (d) 70 percent, and (e) 80 percent CNR degradation, each generated by scan-specific Rician noise. For each row, panels show: (i) the bright-blood phantom cross-section with red arrows indicating scar regions, (ii) a zoomed LA view of (i), (iii) the corresponding DarkSPARC reconstruction, and (iv) a zoomed LA view of (iii). With increasing degradation, bright-blood images exhibit rapid loss of scar–blood contrast and increasingly ambiguous or invisible scar, particularly within thin LA wall segments. DarkSPARC maintains strong dark-blood suppression, reduced background noise, and stable visualization of LA scar across a wide range of CNR loss.

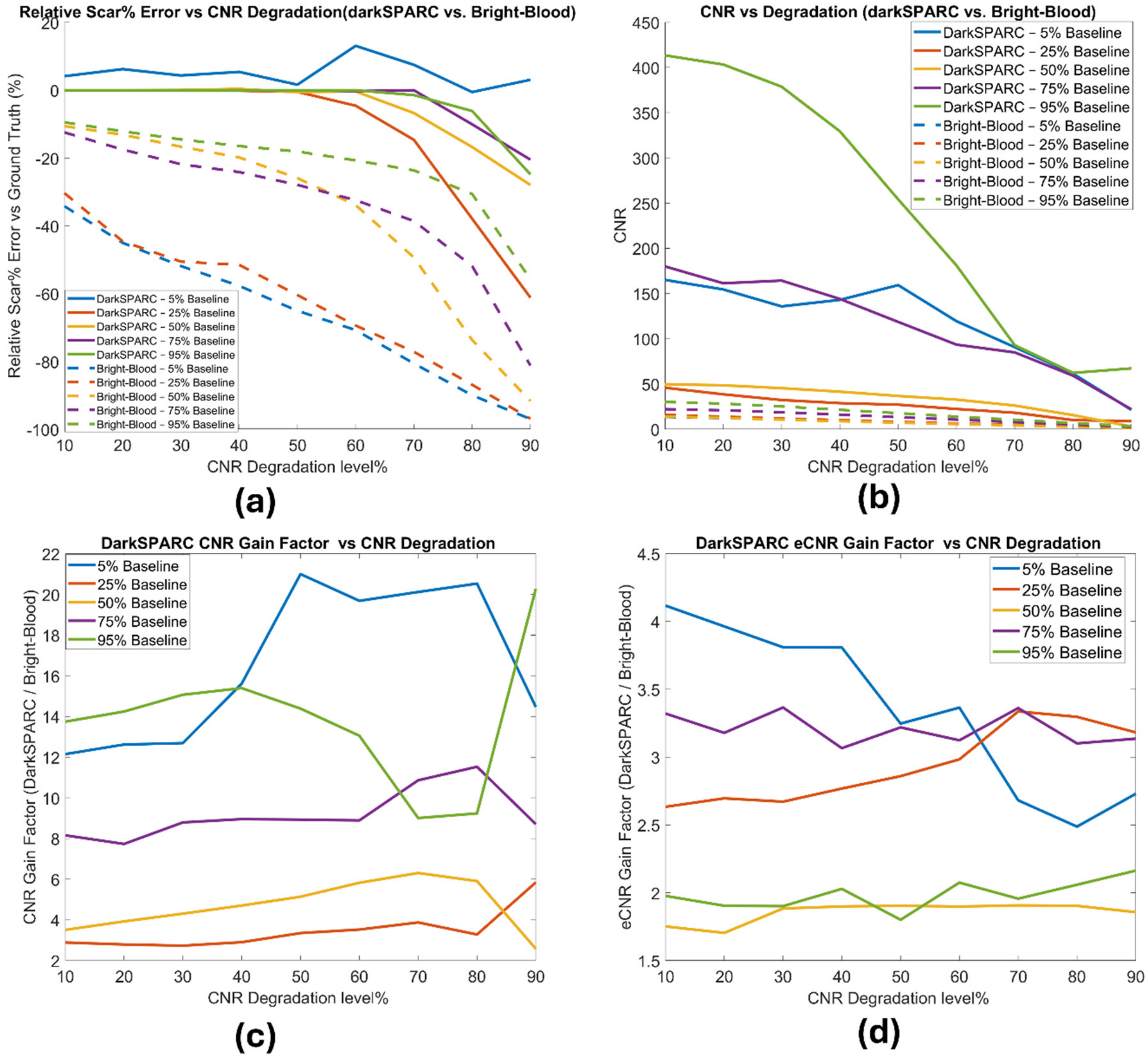

**Figure 5.** Quantitative phantom performance: scar error and CNR/eCNR gains with DarkSPARC across the 5 baseline phantoms. For each of the five baseline scar–pool CNR levels, phantom experiments were performed across nine CNR degradation levels (10 percent to 90 percent CNR loss in 10-percent increments), and results are shown for all phantoms. (a) Relative Scar% error versus ground truth as a function of CNR degradation for bright-blood (dashed) and DarkSPARC (solid) demonstrates that bright-blood increasingly underestimates scar burden (large negative errors) even at mild (≈10 percent) degradation, whereas DarkSPARC errors remain close to zero until very severe degradation. (b) Scar–pool CNR versus degradation shows that DarkSPARC achieves consistently higher, often multi-fold, CNR compared with bright-blood across all five baselines and all degradation levels. (c) CNR gain factor (DarkSPARC / bright-blood) and (d) eCNR gain factor (DarkSPARC / bright-blood) quantify improvements in contrast and effective contrast, respectively, with robust >1-fold gains, frequently several-fold, across the full CNR spectrum and all baselines. Together, these results indicate that DarkSPARC both substantially improves CNR/SNR/eCNR over standard bright-blood LGE and enable more accurate scar quantification with near zero error even under high CNR degradation.

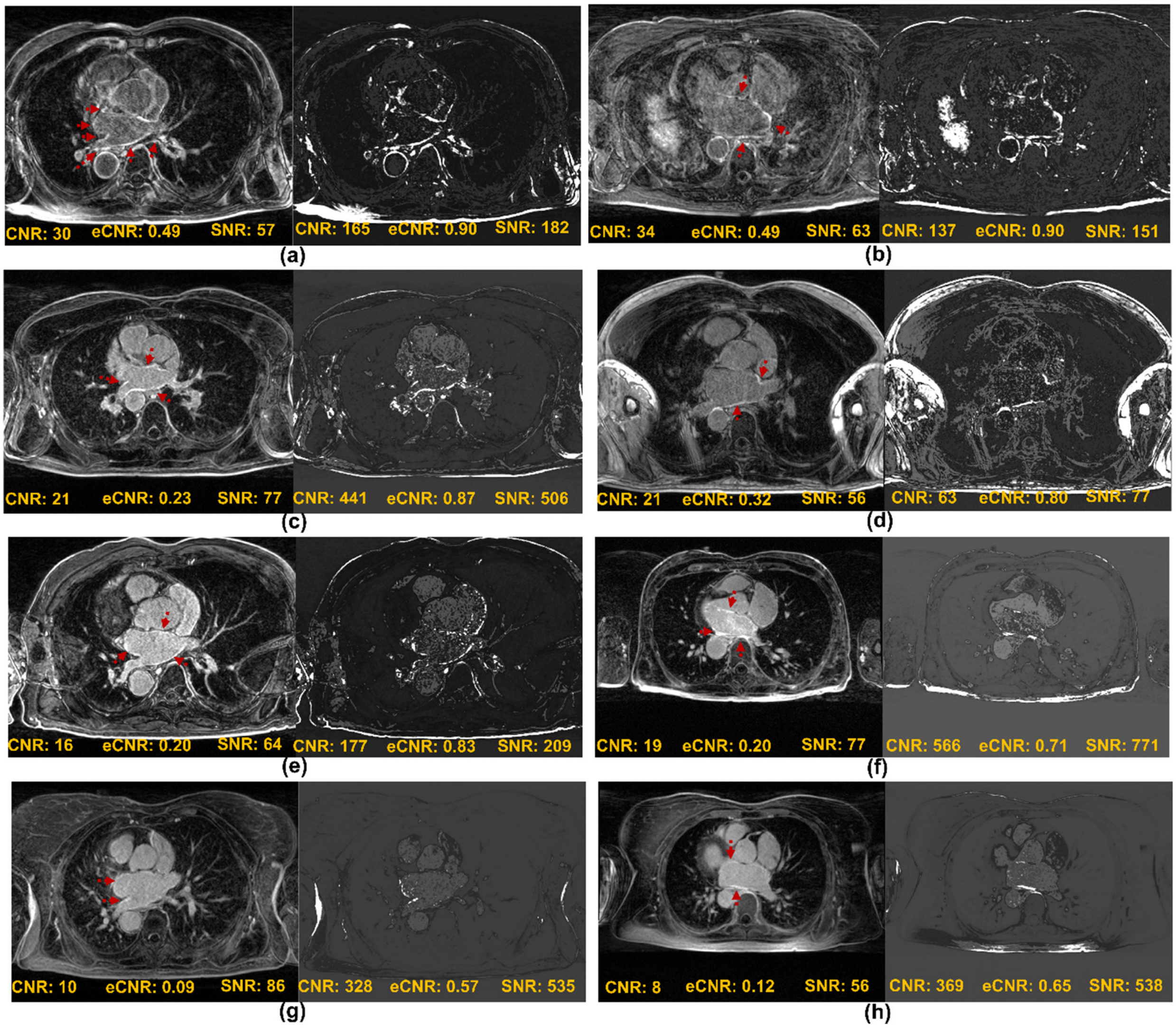

**Figure 6.** Representative in vivo post-ablation examples across the full range of baseline bright-blood CNR. Eight post-ablation patients not used for phantom construction are shown as bright-blood (BB) LGE (left in each pair) and corresponding DarkSPARC reconstructions (right in each pair). Rows correspond to quartiles of baseline BB LA Scar–Pool CNR, from highest to lowest: (a,b) two cases from the highest quartile (75th–100th percentile), (c,d) two cases from the third quartile (50th–75th percentile), (e,f) two cases from the second quartile (25th–50th percentile), and (g,h) two cases from the lowest quartile (0–25th percentile). Arrows on the BB images highlight regions of LA wall hyperenhancement consistent with potential post-ablation scar where low CNR and residual blood-pool signal make scar difficult to appreciate; the same anatomical locations on DarkSPARC images show superior blood suppression, and clearer scar delineation while suppressing background and noise. For each image, LA Scar–Pool CNR, eCNR, and SNR are listed for BB and DarkSPARC, illustrating consistent multi-fold gains in CNR/eCNR and SNR across the entire spectrum of baseline BB CNR.

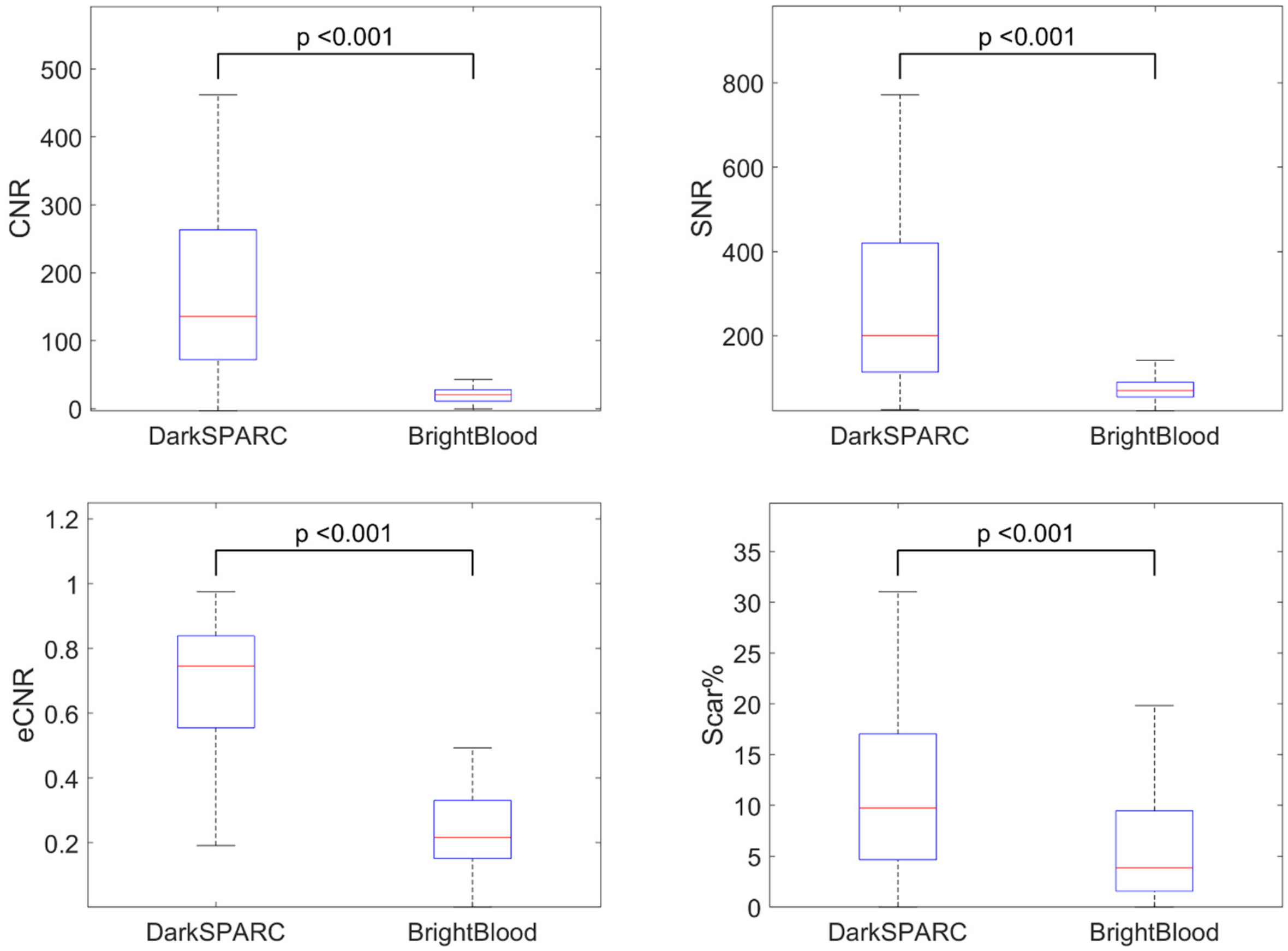

**Figure 7.** Quantitative in vivo performance of DarkSPARC versus bright-blood LGE in 60 post-ablation patients. Boxplots summarize LA Scar–Pool CNR, SNR, eCNR, and LA Scar% for bright-blood (BB) and DarkSPARC reconstructions across all 60 subjects. DarkSPARC significantly increases CNR (approximately 7-fold median gain), SNR (≈3-fold gain), and eCNR (≈3–4-fold gain) compared with BB (all $p < 0.001$, Wilcoxon signed-rank test). LA Scar% (using a $\mu+3.3\sigma$ intensity threshold) is also significantly higher with DarkSPARC than with BB ($p < 0.001$), consistent with phantom experiments showing that bright-blood systematically underestimates true scar burden at low CNR. Together, these in vivo results confirm that DarkSPARC substantially improves blood suppression and scar conspicuity while revealing a larger, more physiologically plausible scar burden compared with conventional bright-blood LGE.